\documentstyle[pra,aps,twocolumn,epsf]{revtex}

\begin{document}
\title{Two-photon interference with thermal light}
\author{Giuliano Scarcelli, Alejandra Valencia, and Yanhua
Shih} \address{Department of Physics, University of Maryland,
Baltimore County, Baltimore, Maryland 21250}\maketitle

\begin{abstract}
The study of entangled states has greatly improved the basic
understanding about two-photon interferometry.  Two-photon
interference is not the interference of two photons but the result
of superposition among indistinguishable two-photon amplitudes.
The concept of two-photon amplitude, however, has generally been
restricted to the case of entangled photons. In this letter we
report an experimental study that may extend this concept to the
general case of independent photons. The experiment also shows
interesting practical applications regarding the possibility of
obtaining high resolution interference patterns with thermal
sources.
\end{abstract}

\vspace{1cm} The superposition principle is probably the most
mysterious and fascinating concept of the theory of quantum
mechanics \cite{feynman}. In  Young's double-slit experiment, a
light quantum has two indistinguishable alternative amplitudes
that result in a photo-electron event at space-time point
($\mathbf{r}, t$).  The superposition of the two indistinguishable
amplitudes produces the interference of the light quantum itself
\cite{Dirac}. Quantum theory may never identify through which slit
(or both slits) the light quantum passed, however, it accurately
predicts the counting rate as a function of the relative delay
between the two amplitudes.

The experimental observations of Hanbury-Brown and Twiss
introduced the concept of second order coherence \cite{hanbury}.
Quantum theory of the second order interferometry describes the
physical process of a joint photo-electron event at space-time
points ($\mathbf{r_{1}}, t_{1}$) and ($\mathbf{r_{2}}, t_{2}$),
produced by two light quanta with distinguishable and/or
indistinguishable alternative amplitudes \cite{fano}. Two-photon
optics is a complex subject involving optical coherence, photon
statistics, and, the nonlocal physics associated with the
Einstein-Podolsky-Rosen two-particle system. Besides probing the
fundamental issues of quantum theory, the massive study of
entangled states, especially the experimental and theoretical
research on the entangled two-photon state of Spontaneous
Parametric Down Conversion (SPDC) has provided great insights on
two-photon interferometry \cite{zeilinger}. In particular, we have
a better understanding of the troubling statement of Dirac: ``Each
photon interferes only with itself. Interference between two
different photons never occurs."  The question whether two
individual photons can or cannot interfere with each other has
been answered experimentally based on the study of two-photon
interferomerty of SPDC: two-photon interference cannot simply be
described in terms of interference of two independent photons but
must be envisioned as an actual two-photon phenomenon in which the
indistinguishable alternatives are two-photon amplitudes
contributing to the final joint photo-electron events
\cite{pittman,yoonho,mandel}.  The concept of two-photon amplitude
is somehow troubling, probably because of the nonlocality that it
implies, and so, if accepted, it has generally been considered
peculiar of entangled photons.

In this letter we wish to report an experimental study that may
extend the concept of two-photon interference as the result of
superposition of indistinguishable two-photon amplitudes to the
general case of two independent photons. The result is intriguing
also in a practical sense because $N$-photon interferometry with
entangled states has been proven to represent a great potential
for imaging and metrological
applications\cite{dowlingmetrolo,zeilingerfour}. In particular it
has been proposed \cite{dowling,scully} and experimentally shown
with two-photon entangled states from SPDC \cite{milena} that it
is possible to do quantum lithography beyond the classical
diffraction limit. In this experiment we simulated the experiment
of D'Angelo et al. \cite{milena} with a pseudo-thermal source of
light \cite{thermaltheory}.  The experiment involves the
measurement of second order interference of pseudo-thermal light
through a standard Young's double slit interferometer.  In order
to provide a clear physical picture of the phenomenon, we will
discuss similar and different aspects between the thermal state
and the entangled state of SPDC in this regard. It must be noted
that a similar source, with a similar setup has been used for
different purposes in a historical experiment\cite{haner}.

The experimental setup is schematically shown in Fig.~\ref{setup}:
the pseudo-thermal light source (or entangled two-photon light
source of SPDC for comparison) illuminates a double slit of
slit-width $a$ with slit-distance $d$;  the joint photo-detection
occurs in the far field plane with two photon counting detectors.

Let's start by noticing that in both cases of pseudo-thermal light
and SPDC radiation there is no observable first order interference
effect in this experimental setup.  The absence of the first order
interference precludes the possibility of the second order
interference to be a consequence of interference of the first
order or the result of ``partial coherence" between the fields at
slit A and slit B. In this particular experiment any observable
interference is not the result of \textit{each photon interfering
with itself}.

The quantity that governs the probability of joint photodetection
and therefore the rate of coincidence counts is the second order
Glauber correlation function \cite{glauber}:
\begin{eqnarray}\label{G2}
G^{(2)}(t_{1},r_{1}; t_{2},r_{2}) \equiv Tr[\hat{\rho}
E_{1}^{(-)}(t_{1}, r_{1})E_{2}^{(-)}(t_{2}, r_{2}) \nonumber \\
E_{2}^{(+)}(t_{2}, r_{2})E_{1}^{(+)}(t_{1}, r_{1}) ]
\end{eqnarray}
where $E_{1,2}^{(\pm )}(r_{j}, t_{j})$, $j=1,2$, are
positive-frequency and the negative-frequency components of the
field at detectors $D_{1}$ and $D_{2}$, and $\hat{\rho}$
represents the density matrix of the quantum state under
consideration. The field operators, in both thermal light and SPDC
cases, can be written as the superposition of earlier fields at
slit A and B:
\begin{eqnarray} \label{efield}
E_{1}^{(+)}(r_{1},t_{1}) & = & E_{A}^{(+)}(r_A,
t_{1}-\frac{r_{A1}}{c})+ E_{B}^{(+)}(r_B, t_{1}-\frac{r_{B1}}{c})\\
\nonumber E_{2}^{(+)}(r_{1},t_{1}) & = & E_{A}^{(+)}(r_A,
t_{2}-\frac{r_{A2}}{c})+ E_{B}^{(+)}(r_B, t_{2}-\frac{r_{B2}}{c})
\end{eqnarray}
with $r_{Aj}$ ($r_{Bj}$) defining the optical path length from
slit $A$ ($B$) to the $j^{th}$ detector. If $G^{(2)}$ is different
in the case of thermal light and SPDC, the difference must come
from the intrinsic property of the light, as expected.

Let's first briefly review the known physics behind the experiment
that uses SPDC as the light source \cite{milena}. SPDC is a
nonlinear process in which an entangled pair of photons, signal
and idler, are simultaneously created. Therefore a joint
photodetection is almost always the result of the detection of the
signal-idler pair.  In ref. \cite{milena}, the double slit was
placed very close to the SPDC crystal so that the source is
divided into two regions: upper slit ($A$) and lower slit ($B$).
Due to the entangled nature, a signal-idler pair is generated
either from slit $A$ or slit $B$, but never from different ones.
It is very intuitive, then, to write the two-photon state that
would lead to a joint detection measurement in the following way:
\begin{equation}\label{spdcstate}
|\Psi \rangle \simeq [a^{\dag}_{s}a^{\dag}_{i}e^{i\phi_{A}}+
b^{\dag}_{s}b^{\dag}_{i}e^{i\phi_{B}}]| 0\rangle
\end{equation}
here $a^{\dag}$ and $b^{\dag}$ stand for the photon creation
operators at the upper and lower slit respectively, and $\phi_{A}$
and $\phi_{B}$ are the phases of the two-photon modes in
correspondence to the upper slit ($A$) and the lower slit ($B$).
The spatial coherence of the pump beam of the SPDC justifies the
assumption: $\phi_{A}-\phi_{B}=constant$. Under these conditions
the expected second order correlation function is calculated as:
\begin{eqnarray} \label{coinspdc-1}
G^{(2)}  & = & \mid e^{i k (r_{A1}+r_{A2})}+ e^{i k
(r_{B1}+r_{B2})}\mid ^{2}\\ \nonumber & \propto &
1+cos[k(r_{A1}+r_{A2}-r_{B1}-r_{B2})]
\end{eqnarray}
In the far field zone, taking into account the finite size of the
slits, the final interference-diffraction pattern is expected as:
\begin{eqnarray} \label{coinspdc}
G^{(2)} \propto sinc^{2}[\frac{\pi a (x_{1}+x_{2})}{\lambda z
}]cos^{2}[\frac{\pi d (x_{1}+x_{2})}{\lambda z}]
\end{eqnarray}
where $x_{1}$ and $x_{2}$ are the horizontal displacement of
$D_{1}$ and $D_{2}$, respectively, and $z$ is the common distance
from  the detectors to the double slit. Besides opening the road
towards quantum lithography,  this experiment clearly demonstrated
that the second order interference is the result of the
superposition between the ``upper-upper" ($A \rightarrow D_{1}$
with $A \rightarrow D_{2}$) and the ``lower-lower" ($B \rightarrow
D_{1}$ with $B \rightarrow D_{2}$) two-photon amplitudes.

In the present experiment, we substituted the SPDC light with a
pseudo-thermal source \cite{martienssen}.  The source is basically
composed by a He-Ne laser beam focused on a rotating ground glass
diffuser disk: the radiation is randomly scattered in all possible
directions. It has been shown theoretically and experimentally
\cite{arecchibook} that the scattered radiation has the same
statistical and optical properties as standard thermal sources.
During the experiment, the intensity of the light was operated in
a very low counting rate regime to achieve the condition in which
only two photons were present in the setup within the joint
detection time window. It is reasonable then to restrict our
analysis at the level of two photons. Let's use a simple physical
model for the process: it can be shown that one possible basis of
the physical state space that describes the two-photon system may
be composed by the following three normalized states \cite{cohen}:
\begin{eqnarray}\label{basis}
|\alpha \rangle & = & a^{\dag}_{k}a^{\dag}_{k'} |0 \rangle;\\
\nonumber | \beta \rangle & = & b^{\dag}_{k}b^{\dag}_{k'}|0
\rangle;\\ \nonumber |\gamma \rangle & = & \frac{1}{\sqrt{2}}(
a^{\dag}_{k}b^{\dag}_{k'}+b^{\dag}_{k}a^{\dag}_{k'}) |0\rangle.
\end{eqnarray}
Here $a^{\dag}$ and $b^{\dag}$ stand for creation operators of the
photons generated at the upper and lower slit respectively; while
$k$, and $k'$ corresponds to the modes of the radiation leading to
detectors $D_{1}$ and $D_{2}$ respectively. Notice that the three
basis vectors correspond to the three intuitive alternatives of
joint photodetection, i.e. ($\alpha$) the two photons both come
from the upper slit $A$; ($\beta$) both come from the lower slit
$B$; or ($\gamma$) one comes from $A$ and the other from $B$.  It
is important to emphasize that ($\gamma$) will lead to the
interference feature. We will treat our system as a statistical
mixture of the three basis vectors:
\begin{equation}\label{combination}
\hat{\rho}= |\alpha|^{2} |\alpha \rangle \langle \alpha | +| \beta
|^{2} |\beta \rangle \langle \beta | + |\gamma |^{2} |\gamma
\rangle \langle \gamma |
\end{equation}
where $|\alpha|^{2}$, $|\beta|^{2}$, and $|\gamma|^{2}$, are the
probabilities of having the system in one of the basis vectors. In
the thermal light case, the three probabilities are equal ($1/3$).
Thus, the second order correlation function can be expressed as
follows:
\begin{eqnarray}\label{G2dens1}
 G^{(2)} \propto \langle \alpha
|E_{1}^{(-)}E_{2}^{(-)}E_{2}^{(+)}E_{1}^{(+)}|\alpha \rangle+\\
\nonumber \langle \beta
|E_{1}^{(-)}E_{2}^{(-)}E_{2}^{(+)}E_{1}^{(+)}|\beta \rangle+\\
\nonumber \langle \gamma
|E_{1}^{(-)}E_{2}^{(-)}E_{2}^{(+)}E_{1}^{(+)}|\gamma \rangle.
\end{eqnarray}
Substituting the field operators of Eq. (\ref{efield}) into Eq.
(\ref{G2dens1}) we obtain:
\begin{eqnarray} \label{coin3}
G^{(2)} \propto |e^{i k (r_{A1}+r_{A2})}|^{2}+
|e^{i k (r_{B1}+r_{B2})}|^{2}+ \\
\nonumber  \frac{1}{2}| e^{i k (r_{A1}+r_{B2})}+ e^{i k
(r_{B1}+r_{A2})}| ^{2}.
\end{eqnarray}
The superposition of the indistinguishable two-photon amplitudes
``upper-lower" ($A \rightarrow D_{1}$ with $B \rightarrow D_{2}$)
and ``lower-upper" ($A \rightarrow D_{2}$ with $B \rightarrow
D_{1}$) are responsible for the interference. In the far field
zone and considering the finite size of the slits, the
interference-diffraction pattern is thus:
\begin{eqnarray} \label{resultfar}
G^{(2)} \propto 1+ sinc^{2}(\frac{ \pi a (x_{1}-x_{2})}{\lambda z}
)cos^{2}(\frac{\pi d (x_{1}-x_{2})}{\lambda z}).
\end{eqnarray}
Comparing with the SPDC case of (Eq.~\ref{coinspdc}), we obtain a
similar interference-diffraction pattern, hinting to the similar
two-photon physics behind the two effects.

In the actual experiment we had an attenuated He-Ne laser beam
impinging on a double slit, $10cm$ after the slit we put a
converging lens ($f=25mm$) and placed the rotating ground glass
disk at $33.5mm$ from the lens. Basically we imaged the double
slit onto the ground glass in order to produce an effective double
slit illumination ($a=0.043mm$ and $d=0.135mm$) on the disk, that
is our source, as indicated in Fig.~\ref{setup}. The radiation
scattered by the ground glass was then divided by a beam splitter
and sent to two horizontally displaceable fibers, connected to
single photon counting modules.

Fig.~\ref{coin} reports the measured two-photon
interference-diffraction pattern. The solid line represents the
theoretical fit using Eq.~\ref{resultfar}. It is interesting to
see that, similarly to the SPDC case, the interference-diffraction
pattern is twice as narrow as the standard
interference-diffraction pattern of He-Ne light (shown in the
lower plot of Fig.~\ref{coin} for comparison) and with
interference modulation twice as large as the standard pattern, as
if it was produced by a source of light with half the wavelength
of the He-Ne laser. The visibility of the pattern, however, is
about $28 \%$  whereas SPDC allows $100 \%$ visibility.
Fig.~\ref{single} reports the single counts of detector $D_{1}$
and $D_{2}$ when the detectors are scanned in the horizontal
direction.   It is apparent that the single counts are flat over
the entire analyzed range. This demonstrates the absence of any
first order interference phenomena, i.e. no ``partial coherence"
existing in this experiment. In order to ensure that the source we
used was indeed ``thermal", before proceeding to the actual
measurement, we repeated the historical experiment performed by
Arecchi et al \cite{arecchi}. Fig.~\ref{mca} reports the result of
the second order correlation measurements. Basically we plot a
histogram of number of coincidence counts versus the time
difference of the clicks from the two detectors. This measurement
is the principal evidence of thermal light statistics of a
pseudothermal source. This result was also used to ``calibrate"
the coincidence time window. In all the measurements of this
experiment, we set a time window of about $600$ ns around the peak
of Fig.~\ref{mca} and measured the number of coincidences within
that window.

The experimental results therefore confirm the expected
similarities and differences between the thermal light and the
entangled state regarding to the two-photon interference.  As was
shown in the theoretical derivation, the analogy between the
thermal case and the SPDC case is due to the similar physics: the
superposition of the two-photon alternative amplitudes leads to
the interference. However there are two main differences:  (1) the
joint detection counting rate is a function of $x_{1}-x_{2}$ in
the thermal case instead of $x_{1}+x_{2}$. The reason is that in
the SPDC case, the two-photon amplitudes leading to the
interference are the ``upper-upper" ($A \rightarrow D_{1}$ with $A
\rightarrow D_{2}$) and the ``lower-lower" ($B \rightarrow D_{1}$
with $B \rightarrow D_{2}$) alternatives; in the case of thermal
light, instead, the interference is produced by the superposition
of the ``upper-lower" ($A \rightarrow D_{1}$ with $B \rightarrow
D_{2}$) and the ``lower-upper" ($A \rightarrow D_{2}$ with $B
\rightarrow D_{1}$) alternatives; (2) the visibility in the
thermal light case is limited to $1/3$, in fact in the case
analyzed in this experiment the ``upper-upper" and ``lower-lower"
alternatives do not contribute to the interference, but they do
contribute to the constant background of the pattern. In the SPDC
case, instead, due to entanglement, the only existing two-photon
amplitudes are the indistinguishable ``upper-upper" and
``lower-lower" alternatives.

From a practical point of view it is important to notice that with
a pseudo-thermal source we achieved the same doubling in spatial
resolution that was obtained with entangled two-photon states.
Recently it has been shown that with entangled four-photon states
it is possible to achieve resolutions four times better than the
standard limit\cite{zeilingerfour}. However notice that since in
our experiment the source of light does not involve any nonlinear
process (i.e.: the wavelength of the two photons is the same of
the pump, while in SPDC the entangled photons have twice the
wavelength of the laser pump), the increase in spatial resolution
is twice as large as the increase obtained in the analogous (same
$N$, number of entangled photons) case with entangled photons.
Moreover notice that in the historical experiment similar to ours
\cite{haner} the increase in spatial resolution was not obtained,
however the similarities between the two experiments and the fact
that in \cite{haner} the intensity of light was much higher, lead
us to think that with this source of light it can be overcome the
main limitation presented by entangled photon sources at the
moment, i.e.: the low counting rates. The practical applicability
of the method shown here could be precluded by the low visibility
of the pattern. For certain measurements, however, it may be
possible to implement a detection scheme insensitive to the
constant background noise that could restore high visibilities.

In conclusion, we have experimentally studied a second order
interference phenomenon with thermal light. By comparing this
experiment with the analogous one performed with entangled
photons, we have justified the physical interpretation of the
phenomenon in terms of interference between indistinguishable
two-photon amplitudes. Paraphrasing Dirac, we may summarize the
physics as follows: although \textit{each photon did not interfere
with itself} in this experiment, the observed interference is the
result of each \textit{pair} of independent photons interfering
with itself.

The authors would like to thank S. Thanvanthri, J. Wen,
M.D'angelo, D.Hudson and M.H. Rubin for helpful discussions. This
research was supported in part by NSF, ONR and NASA-CASPR program.

\begin{figure}
\caption{Sketch of the experimental setup. An attenuated He-Ne
laser beam illuminates a double slit, $10cm$ after the slit there
is a converging lens ($f=25mm$) and a the rotating ground glass
disk is placed at $33.5mm$ from the lens. Basically the double
slit is imaged onto the ground glass producing an effective double
slit illumination ($a=0.043mm$ and $d=0.135mm$). The radiation
scattered by the ground glass is then divided by a beam splitter
and sent to two horizontally displaceable fibers, connected to
single photon counting modules.} \label{setup}
\end{figure}

\begin{figure}
\caption{(a) Normalized second order interference diffraction
pattern vs position of the detectors. The dots are the
experimental data while the solid line is a theoretical fit from
Eq.~\ref{resultfar}. The actual counting rate corresponds to about
$1000$ coincidence counts per second in the peak. The single
counts are about $45000$ per second in $D_{1}$ and $25000$ per
second in $D_{2}$.(b) Equivalent first-order interference
diffraction pattern.} \label{coin}
\end{figure}

\begin{figure}
\caption{Single detector counts vs positions of the detector
$D_{1}$ (filled circles) and $D_{2}$ (hollow circles). The low
level counting rate shows that the experiment was performed in the
two-photon regime. The flatness of the graphs shows the absence of
first order interference.} \label{single}
\end{figure}

\begin{figure}
 \caption{Histogram of
number of joint detection counts vs time difference of the two
photo-electron events. The size of each channel is $0.3$ ns. The
graph is useful to ``calibrate" the coincidence time window: in
all the measurements of this experiment, the coincidence were
counted in a time window of about $600$ ns around the peak of the
figure while the noise background was verified by shifting the
coincidence window of $4000$ ns towards the region of the
accidental coincidences.} \label{mca}
\end{figure}

\centerline{\epsfxsize=2in \epsffile{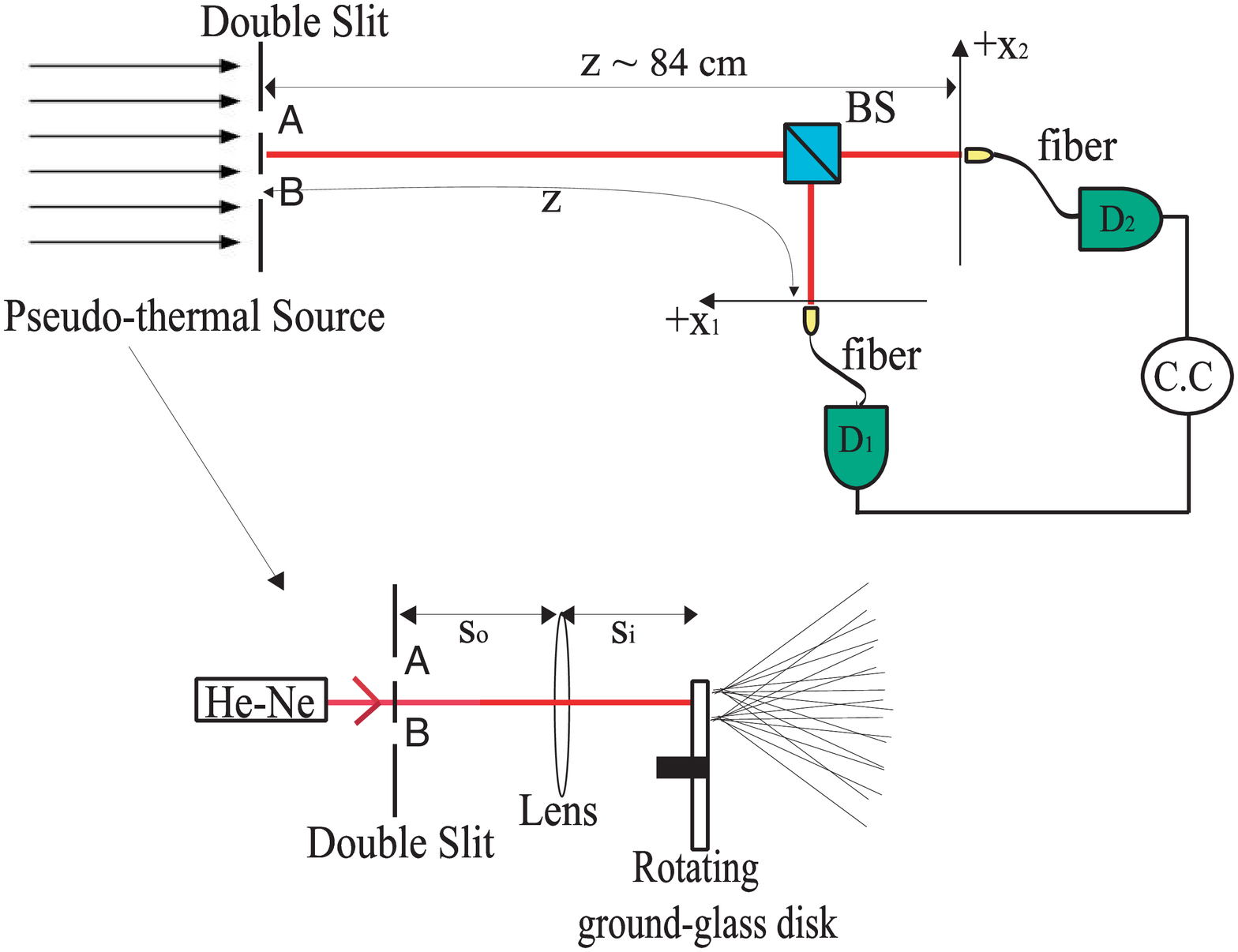}} \vspace{1cm}
Figure \ref{setup}.  Giuliano Scarcelli, Alejandra Valencia, and
Yanhua Shih.

\centerline{\epsfxsize=2.5in \epsffile{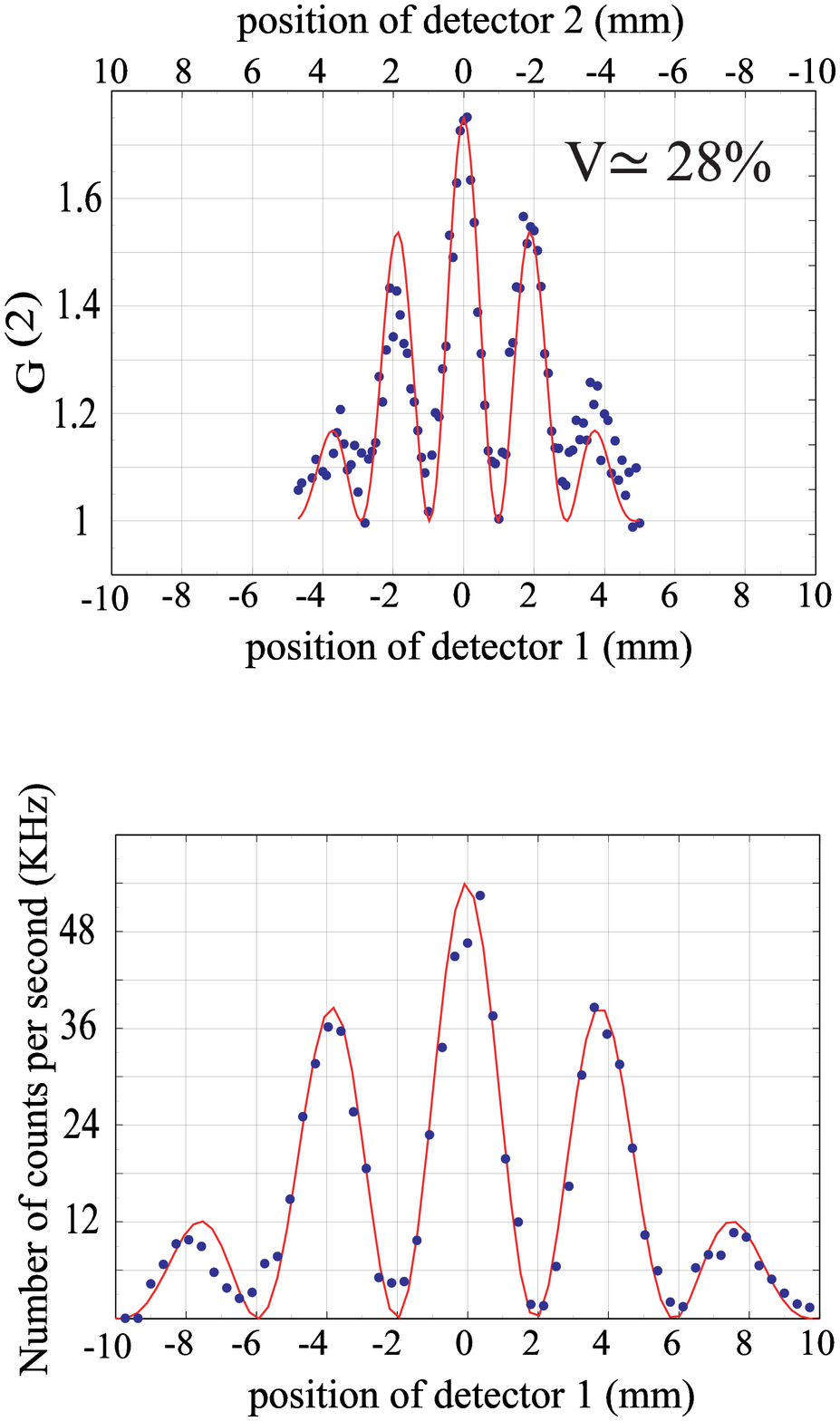}} \vspace{1cm}
Figure \ref{coin}.  Giuliano Scarcelli, Alejandra Valencia, and
Yanhua Shih.

\centerline{\epsfxsize=2.5in \epsffile{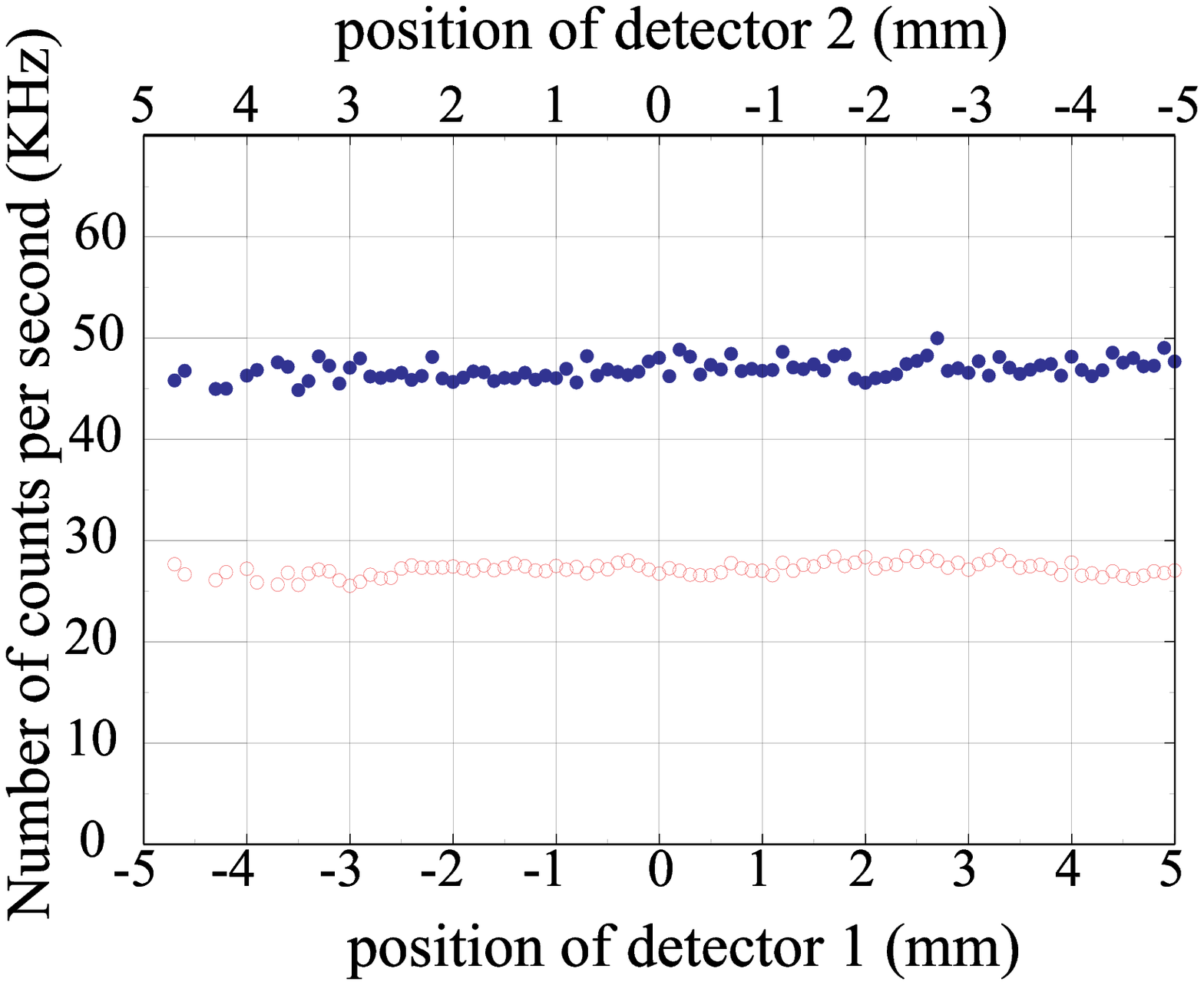}} \vspace{1cm}
Figure \ref{single}.  Giuliano Scarcelli, Alejandra Valencia, and
Yanhua Shih.

\centerline{\epsfxsize=2.5in \epsffile{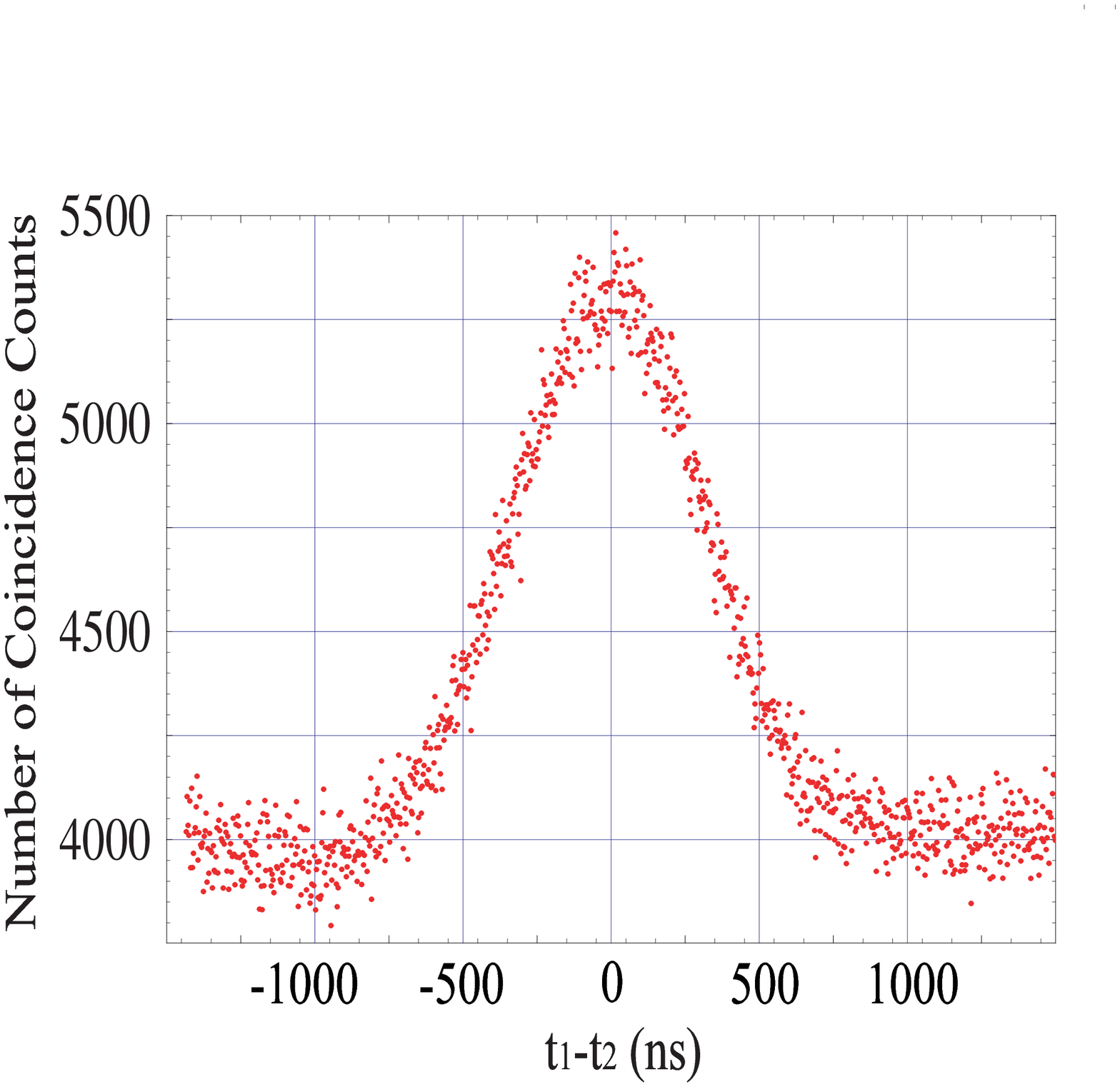}} \vspace{1cm}
Figure \ref{mca}.  Giuliano Scarcelli, Alejandra Valencia, and
Yanhua Shih.

\end{document}